\documentclass[aps,prb,preprint,groupedaddress,amsmath,amssymb, showpacs]{revtex4-1}
\usepackage{epsfig,psfrag}
\usepackage{dcolumn}
\usepackage{bm}
\usepackage{epsfig} 
\usepackage{graphicx}
\usepackage{epstopdf}
\usepackage{xcolor}
\usepackage{array}
\usepackage{import}
\usepackage{siunitx}
\begin{document}
\title{Single photon superradiance and cooperative Lamb shift in an optoelectronic device}

\author{G. Frucci}
\thanks{These authors contributed equally to this work.}
\affiliation{Universit\'e Paris Diderot, Sorbonne Paris Cit\'e, Laboratoire Mat\'eriaux et Ph\'enom\`enes Quantiques, UMR 7162, 75013 Paris, France}
\author{S. Huppert}
\thanks{These authors contributed equally to this work.}
\affiliation{Universit\'e Paris Diderot, Sorbonne Paris Cit\'e, Laboratoire Mat\'eriaux et Ph\'enom\`enes Quantiques, UMR 7162, 75013 Paris, France}
\author{A. Vasanelli}
\email{Electronic address: angela.vasanelli@univ-paris-diderot.fr}
\affiliation{Universit\'e Paris Diderot, Sorbonne Paris Cit\'e, Laboratoire Mat\'eriaux et Ph\'enom\`enes Quantiques, UMR 7162, 75013 Paris, France}
\author{B. Dailly}
\affiliation{Universit\'e Paris Diderot, Sorbonne Paris Cit\'e, Laboratoire Mat\'eriaux et Ph\'enom\`enes Quantiques, UMR 7162, 75013 Paris, France}
\author{Y. Todorov}
\affiliation{Universit\'e Paris Diderot, Sorbonne Paris Cit\'e, Laboratoire Mat\'eriaux et Ph\'enom\`enes Quantiques, UMR 7162, 75013 Paris, France}
\author{G. Beaudoin}
\affiliation{Laboratoire de Photonique et de Nanostructures, CNRS, 91460 Marcoussis, France}
\author{I. Sagnes}
\affiliation{Laboratoire de Photonique et de Nanostructures, CNRS, 91460 Marcoussis, France}
\author{C. Sirtori}
\affiliation{Universit\'e Paris Diderot, Sorbonne Paris Cit\'e, Laboratoire Mat\'eriaux et Ph\'enom\`enes Quantiques, UMR 7162, 75013 Paris, France}

\begin{abstract}
Single photon superradiance is a strong enhancement of spontaneous emission appearing when a single excitation is shared between a large number of two-level systems. This enhanced rate can be accompanied by a shift of the emission frequency, the cooperative Lamb shift, issued from the exchange of virtual photons between the emitters. In this work we present a semiconductor optoelectronic device allowing the observation of these two phenomena at room temperature. We demonstrate experimentally and theoretically that plasma oscillations in spatially separated quantum wells interact through real and virtual photon exchange. This gives rise to a superradiant mode displaying a large cooperative Lamb shift.
\end{abstract}

\maketitle

Superradiance is one of the many fascinating phenomena predicted by quantum electrodynamics that have first been experimentally demonstrated in atomic systems~\cite{skribanowitz, goban} and more recently in condensed matter systems like semiconductor quantum dots~\cite{scheibner}, superconducting q-bits~\cite{vanLoo}, cyclotron transitions~\cite{zhang} and plasma oscillations in quantum wells (QWs)~\cite{laurent_PRL}. It occurs when a dense collection of $N$ identical two-level emitters are phased via the exchange of photons, giving rise to enhanced light-matter interaction, hence to a faster emission rate ~\cite{dickePR1954, haroche_gross_PRep1982}. Superradiance can be obtained by preparing the emitters in different ways: a well known procedure is to promote all of them in the excited state and observe their coherent decay through successive emission of $N$ photons into free space~\cite{haroche_gross_PRep1982}. Of great interest is also the opposite regime where the ensemble interacts with one photon only and therefore all of the atoms, but one, are in the ground state. In this case the quantum superposition of all possible single emitter excitations produces a symmetric state that decays radiatively with a rate $N$ times larger than that of the individual oscillators. This phenomenon is called single photon superradiance~\cite{scully_science2009} and was first predicted by Dicke~\cite{dickePR1954}, whose model describes the phasing of the emitters by the exchange of real photons. Yet, single photon superradiance is also associated with another collective effect that arises from virtual photon exchanges triggered by the vacuum fluctuations of the electromagnetic field. This phenomenon, known as cooperative Lamb shift~\cite{scully_science2010, scully_PRL2010, FriedbergPR2010_lamb}, renormalizes the emission frequency, and was only recently evidenced experimentally in atomic systems~\cite{rohlsberger_science2010, Keaveney_PRL2012, Meir}.

In this work, we show that single photon superradiance and cooperative Lamb shift can be engineered in a semiconductor device by coupling spatially separated plasma resonances arising from the collective motion of confined electrons in QWs. These resonances are associated with a giant dipole along the growth direction $z$. They have no mutual Coulomb coupling and interact only through absorption and re-emission of real and virtual free space photons. They thus behave as a collection of \textit{macro-atoms} located on different positions along $z$. 
Our device is therefore very valuable to simulate the low excitation regime of quantum electrodynamics in a solid state system.

\begin{figure}[htbp] 
\centering  
\includegraphics[width=0.7\linewidth]{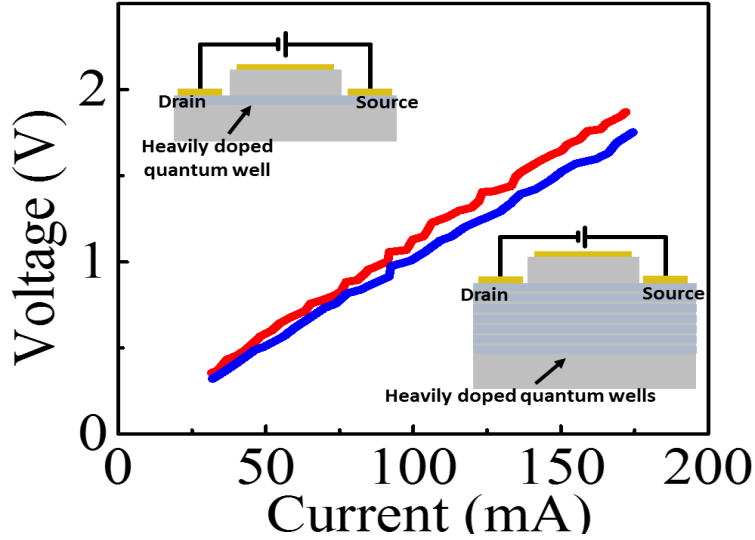}
\caption{Schematic representation of the two samples used in this study with their current-voltage characteristic: the single QW sample (top left-hand corner, red lines) and 
the multiple QW sample (bottom right-hand corner, blue lines). The electrical resistance is $\approx 10 ~\Omega$.}
\label{sample_scheme}
\end{figure}

The two samples used in this study are based on GaInAs/AlInAs highly doped QWs grown by metal organic chemical vapor deposition on an InP substrate. 
The first one (SQW) consists of a single 45 nm GaInAs layer, n-doped with a surface density $N_s=7.5 \times 10^{13}$~cm$^{-2}$, sandwiched between two AlInAs barriers. 
The second sample (MQW) is designed such that six QWs, identical to that of SQW sample, are distributed within one wavelength and separated from one another by a sufficiently thick barrier to avoid tunneling. 

Both samples are processed into field effect transistor-like structures (Insets of Fig.1), consisting of two ohmic contacts for source -- drain current injection and a top mirror. For the MQW sample, we expressly connected electrically only one QW by depositing ohmic contacts directly on the first GaInAs layer. As a consequence in the MQW device electrons located in different wells only interact via the exchange of free space photons. Figure~\ref{sample_scheme} shows that SQW (red lines) and MQW (blue lines) devices display the same source -- drain voltage -- current characteristics. 

In previous work~\cite{laurent_PRL} we demonstrated that the plasma resonance of a highly doped QW, called {\textit{multisubband plasmon}}~\cite{delteilPRL2012plasmons}, can be thermally excited by applying a current through a source -- drain contact. The multisubband plasmons, issued from Coulomb interaction among electronic transitions within the conduction band of the well, superradiantly decay into free space, with a rate proportional to the electronic density in the QW, $N_s$. The emission spectrum measured at a given internal angle $\theta$ presents a unique peak, whose linewidth contains a non-radiative contribution $\gamma$ and a radiative one $\Gamma(\theta)$ given by: 
\begin{equation}
\Gamma (\theta)=\Gamma_0 \sin^2 {\theta} /\cos{\theta} \propto N_s \sin^2 {\theta} /\cos{\theta}
\label{gamma}
\end{equation} 
where $\Gamma_0$ depends on $N_s$ and on the confining potential~\cite{huppert_PRB2016} (see supplementary material). The dependence on $\sin^2 {\theta}$ accounts for the fact that the plasmon collective dipole is oriented along the growth direction $z$ of the QW. Due to the lack of wavevector conservation along $z$, the plasmon interacts with a one-dimensional density of photon states, resulting in the $1/\cos{\theta}$ factor. The radiative broadening $\Gamma(\theta)$ characterizes the strength of the light -- matter interaction. By varying the angle we can therefore explore very different regimes of interaction as $\Gamma(\theta)$ varies from zero to a divergence. 
\begin{figure}[ht] 
\centering  
\includegraphics[width=0.8\linewidth]{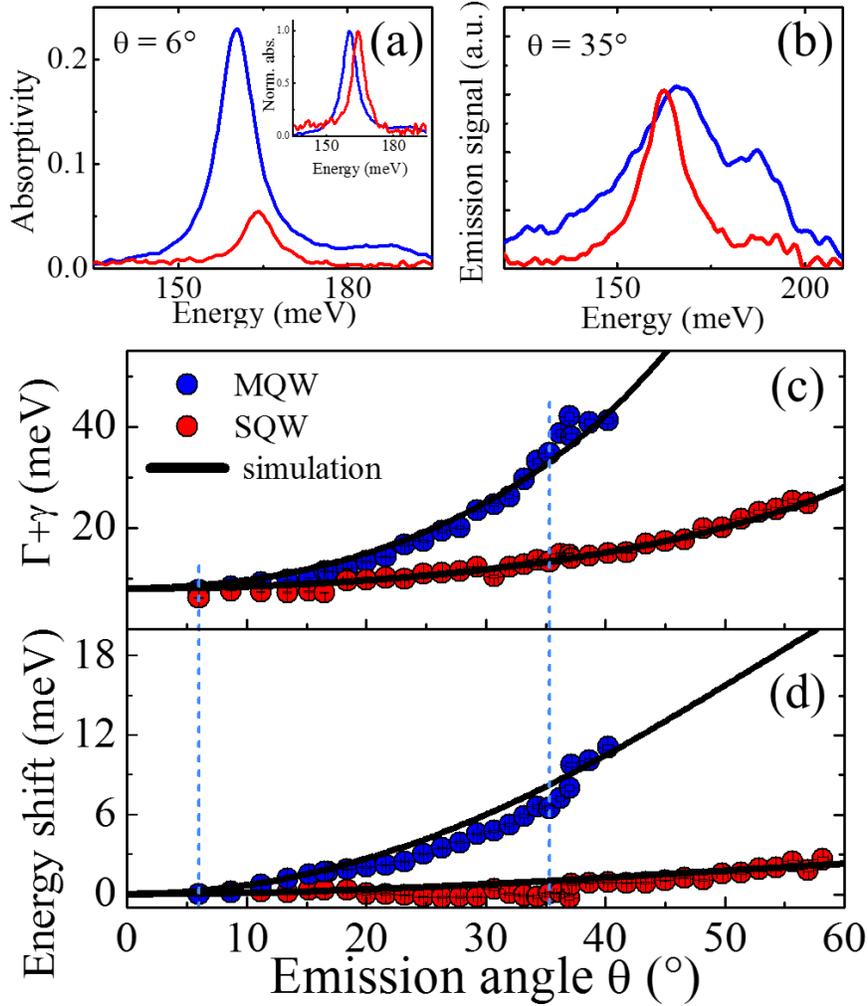}
\caption{Summary of the absorption and thermal emission measurements. Red symbols (lines) refer to SQW and blue symbols (lines) to MQW samples. (a) Absorptivity spectra at $\theta=6^\circ$.  (b) Emission 
spectra at $\theta=35^\circ$. (c) Linewidth of the main plasmon peak, compared with the values calculated using eq.~\ref{gamma} (black line).  (d) Energy shift of the main plasmon peak with respect to 6$^\circ$ as a function of $\theta$ and its comparison to the theoretical expression of the cooperative Lamb 
shift (black line). The vertical dashed lines indicate the two angles corresponding to the graphs (a) and (b).}
\label{broadening}
\end{figure}

At low values of $\theta$, when the coupling is weak, the linewidth is dominated by non-radiative effects. We have experimentally studied this regime between $6^\circ$ and $16^\circ$ by performing absorption measurements through the substrate on unprocessed samples, with only a gold mirror on the top surface. 
Figure~\ref{broadening}a presents two spectra measured at $6^\circ$ internal angle on 
SQW (red line) and MQW (blue line) sample. The two normalized spectra are identical (see inset of fig.~\ref{broadening}a) with a plasmon resonance at $\hslash\omega_0 \approx 165$~meV and very similar linewidths (7.6 meV and 
7.1 meV respectively for MQW and SQW). As it can be observed in fig.~\ref{broadening}a, the peak absorptivity of the MQW sample is much bigger than that of SQW. Indeed, in agreement with the outcome of a perturbative description of the interaction~\cite{helm}, the peak absorptivity of the multisubband plasmon mode is 
proportional to the number of QWs effectively coupled with the electromagnetic field $n_{QW}$. Due to the presence of a metallic mirror on the top of the sample, $n_{QW}(\theta)=\sum_n{\vert \cos(q z_n) \vert^2}$. Here $q=\frac{\sqrt{\epsilon_s}}{c}\omega_0 \sin \theta$ is the projection of the photon wavevector along $z$, $\epsilon_s$ is the InP dielectric constant and $z_n$ is the position of the $n$th QW with respect to the metallic mirror. For the MQW device $n_{QW}(6^\circ)\sim 4$. 

When the angle is increased, the radiative broadening $\Gamma(\theta)$ increases and becomes dominant over $\gamma$. Figure~\ref{broadening}b compares two emission spectra measured at $\theta = 35^\circ$ for the SQW (red line) and the 
MQW (blue line) samples (see supplementary material for the technique used to obtain these spectra). Although the two samples are made of identical QWs (one for SQW and six for MQW) their emission spectra have a 
completely different shape. The main multisubband plasmon peak is much broader for MQW sample than for SQW. Furthermore, a second resonance at $\approx 185$~meV, , associated with an 
excited multisubband plasmon mode~\cite{pegolotti_PRB}, is much more apparent in the MQW than in the SQW sample. Finally, the total incandescence signal of MQW device (directly related to absorption by Kirchhoff's law~\cite{huppert_ACS}) is only twice the SQW one. This is a strong evidence that, contrary to the low angle case, light-matter interaction is not perturbative and has to be described by an exact model~\cite{huppert_PRB2016}.

Plasmons are thermally excited by applying a current between source and drain, modulated at a frequency of 10 kHz with a 50\% duty 
cycle~\cite{laurent_PRL} at a fixed electrical power of 400 mW. 
Red (blue) bullets in fig.~\ref{broadening}c present the full width at half the maximum, $\gamma+\Gamma(\theta)$, of the main multisubband plasmon peak, extracted from emission and absorption measurements on the SQW (MQW) device, 
as a function of the internal angle $\theta$. For the SQW device, the data follow very well~\eqref{gamma} (black line), with a rate $\hslash \Gamma_0=13$ meV corresponding to the nominal density of electrons in the QW. The larger broadening of the main emission peak in the MQW device indicates a much faster radiative decay for this sample than in SQW. This arises because multiple photon absorption and re-emission mediate an effective interaction between plasmons located in different QWs. This light-mediated interaction between spatially separated plasmons gives rise to a superradiant mode extending over all the QWs in the structure, which gathers the oscillator strength of all plasmons.  

Figure~\ref{broadening}d presents the measured shift of the main peak position (with respect to the peak energy at $\theta = 6^\circ$) extracted from absorption and emission measurements. 
While the shift is negligible for the SQW sample (2 meV between $0^\circ$ and $55^\circ$), it becomes substantial for MQW. The observed blueshift corresponds to a cooperative Lamb shift arising from virtual photon 
emission and reabsorption processes~\cite{scully_science2010}.
 
\begin{figure}[ht] 
\centering  
\includegraphics[width=0.8\linewidth]{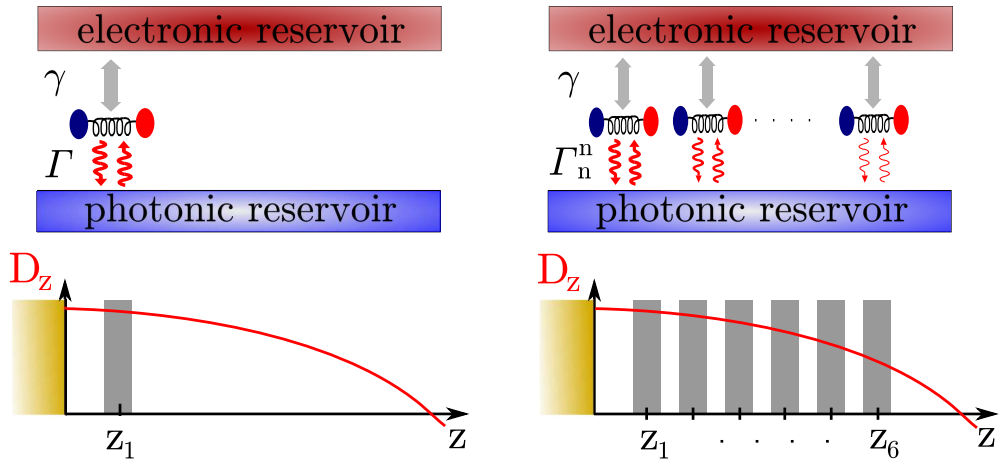}
\caption{Top panels: schematic representation of the system used to model plasmon emission through quantum Langevin equations. Each macro-atom is represented as a dipole, coupled with two reservoirs.
The lower panels represent schematically the variations of the electric displacement field $D_z$ for transverse magnetic radiation in SQW (left) and MQW (right) samples: the field decreases like $\cos(qz)$.}
\label{schema}
\end{figure}

In order to prove our physical interpretation of the experimental observations, we have extended the non-perturbative model developed in previous work~\cite{huppert_ACS, huppert_PRB2016} to the multiple QW case. Our model relies on quantum Langevin equations, describing the dynamics of the plasmon operators, coupled with an electronic and a photonic bath, as schematized on the top panel of fig.~\ref{schema}. The annihilation operator of the main plasmon mode located in the QW of index $n$ 
(at position $z_n$, see lower panels in fig.~\ref{schema}) and characterized by an in-plane wavevector $\mathbf{k}$ is denoted $P_{n,\mathbf k}$.  For simplicity, only the main plasmon mode at energy $\hslash \omega_0$ 
is included in this theoretical discussion (see supplementary materials for the full theoretical method, employed to simulate our experiments). The variations of $P_{n, \mathbf k}$ 
are given by quantum Langevin equations: 
\begin{equation}
\frac{dP_{n,\mathbf k}}{dt}=\left[-i \omega_0 - \frac{\gamma}{2} \right] P_{n,\mathbf k} (t) 
 - \sum_{n'}  \frac{\Gamma_{n}^{n'}(\theta)}{2} \  P_{n',\mathbf k} (t)   +  F_{n,\mathbf k} (t)   \label{langevin}
\end{equation}
In the above, $\omega_0$ and $\gamma$ are considered independent on the QW index $n$, as all QWs are identical. 
The operator $F_{n,\mathbf k}$ is the Langevin force, that arises from the interaction of plasmons with their fluctuating environment. This force is responsible of the thermal excitation of plasmons. 
The rates $\Gamma_{n}^{n'}(\theta)$ characterize the exchanges between plasmons through the bath of free space photons. 
Considering the spatial distribution of the electromagnetic field in the presence of the top mirror, the radiative rates $\Gamma_{n}^{n'}$ can be written as (see supplementary materials for their derivation): 
\begin{eqnarray}
 \text{Re} \left[ \Gamma^{n'}_{n } (\theta) \right]  &=&  \cos(qz_n)\cos(q z_{n'})\  \Gamma (\theta) \label{realGamma_n} \\ 
  \text{Im}\left[\Gamma_{n}^{n'}(\theta)\right]&=& \frac{\sin(q|z_n-z_{n'}|)+\sin(q|z_n+z_{n'}|)}{2} \  \Gamma (\theta)  \label{imagGamma_n}
\end{eqnarray}
The real part of the radiative rates is related to the exchange of real photons, and it determines the radiative broadening, while the imaginary part, 
associated with the exchange of virtual photons, gives rise to the Lamb shift of the emission energy.
In the SQW case (i.e. $n=n'=1$ and $qz\ll1$) the real part of the radiative rate is given by \eqref{gamma}, while its imaginary part, 
the SQW Lamb shift, is negligible (see the black line on fig. \ref{broadening}d). 
In the MQW case, the rates $\Gamma_{n}^{n}(\theta)$, that describe the 
radiative decay of plasmons in each well, depend on the QW index $n$ and their imaginary part cannot be neglected. Furthermore, non-diagonal rates, corresponding to photon-mediated 
exchanges between QWs, give rise to a single superradiant mode in which plasmons of all the different QWs oscillate in phase. 

In order to gain further insight, we have derived the Langevin equation for this superradiant mode issued from the spatially separated plasmons. Due to the variations of the field over the structure thickness, 
the superradiant mode does not correspond exactly to the symmetric combination of all plasmons. Its annihilation operator is instead given by: 
$P_{S, \mathbf k}= n_{QW}(\theta)^{-\frac{1}{2}} \   \sum_n \cos(q z_n) P_{n, \mathbf k}$.
Its dynamics is described by the single operator equation:
\begin{equation}
 \frac{dP_{S,\mathbf k}}{dt}=\left\{-i \Big[\omega_0+ L_{S}(\theta)\Big]- \frac{\gamma}{2}-n_{QW}(\theta) \frac{\Gamma(\theta)}{2} \right\} P_{S,\mathbf k}(t) + F_{S,\mathbf k} (t)   \label{langevin_bright}
\end{equation}
This mode is thus characterized by a superradiant emission rate $n_{QW}(\theta) \Gamma(\theta)$. Indeed, 
experimental data for MQW device in fig.~\ref{broadening}a are well reproduced by~\eqref{gamma} replacing $\Gamma_0$ with $n_{QW}(\theta)\  \Gamma_0$ (black line in fig.~\ref{broadening}a), with $n_{QW}(\theta) \approx 4$ at 
low angle and tends to 6 (the actual number of QWs) when $\theta \rightarrow 90^\circ$. The superradiant mode frequency is also shifted by $L_{S}(\theta)$. This is 
the cooperative Lamb shift that arises from inter-well virtual photon exchanges, that are described by the imaginary part of the rates $\Gamma_{n}^{n'}$. Using \eqref{langevin} we derive an analytical expression 
for $L_S (\theta)$ (see supplementary materials), which tends to increase with the number of QWs, but also depends in an intricate way on the QW positions $z_n$. The calculated Lamb shift is compared with the observed blueshift of the resonance in fig.~\ref{broadening}d (full black line), showing a remarkable agreement for both samples that confirms our physical interpretation.  

\begin{figure}[ht] 
\centering  
\includegraphics[width=0.75\linewidth]{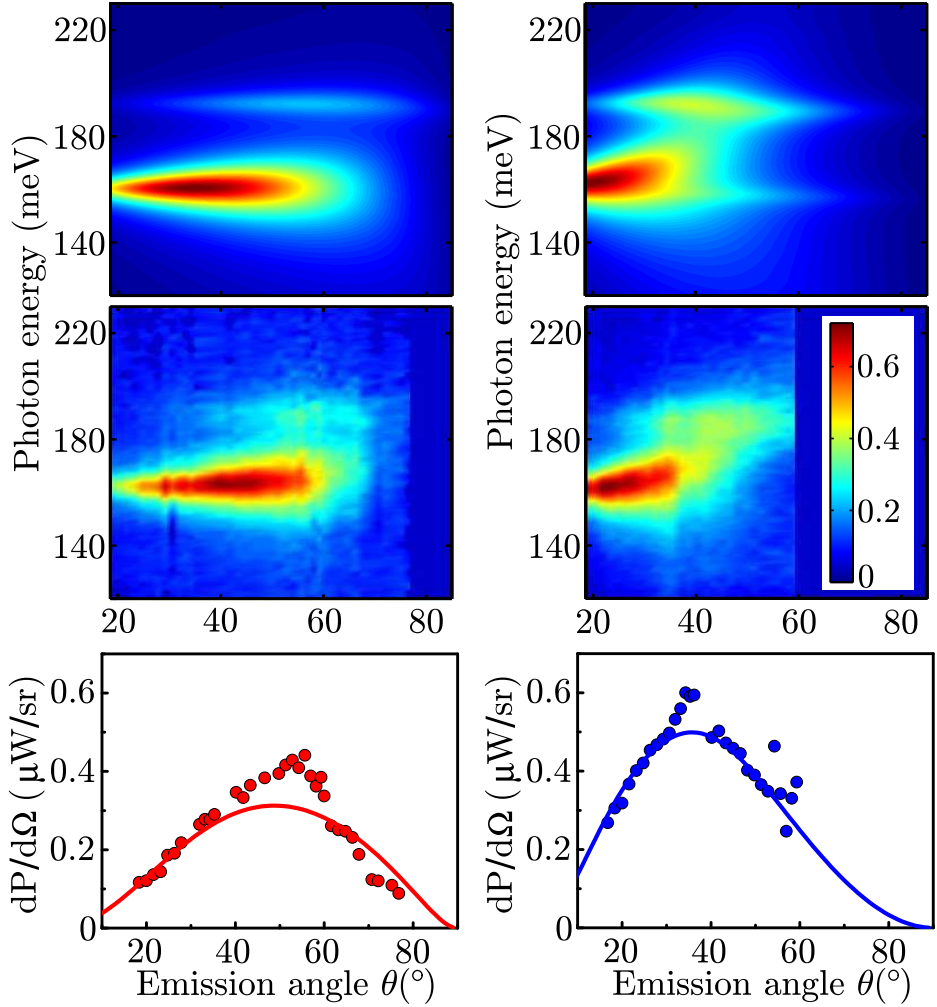}
\caption{Theoretical (top panels) and experimental (middle panels) emission spectra as a function of the emission angle. The emission pattern obtained for SQW (left) and MQW (right) are compared using the same colorscale for 
the optical power. Bottom panels: emitted power per solid angle as a function of the angle for SQW (left) and MQW (right) samples. The colored dots represent the measured value while the solid lines correspond to the results of quantum Langevin model.}
\label{lamb}
\end{figure}
In order to simulate the full emission spectra, we considered an incoherent thermal input at temperature $T$ in the electronic bath, corresponding to a Langevin force:
\begin{equation}
 \langle  F_{n,\mathbf k}^\dag (\omega')  F_{n,\mathbf k'} (\omega) \rangle = 2\pi \gamma \  \delta_{\mathbf k}^{\mathbf k'} \  \frac{\delta(\omega-\omega')}{e^{\frac{\hslash \omega}{kT}}-1}
\end{equation}
For the SQW sample, we have considered that the current flowing in the doped QW induces a temperature increase $\Delta T=70$~K of the electronic bath, with respect to the substrate temperature $T_0=300$~K ($T=T_0+\Delta T$). 
In the MQW device we assume that, although only one QW is electrically contacted, the electronic temperature increases equally in all the QWs during the electrical pulses, due to the small thickness of the sample. The outcomes of our model (including all 
the plasmon modes) are summarized in the top panels of fig.~\ref{lamb}, which present the complete angular and energy behavior of the two devices, compared to the 
experimental results (middle panels). For the SQW device, the Lamb shift is negligible and the emission peak is maximum when the 
critical coupling condition~\cite{huppert_ACS} $\gamma=\Gamma(\theta)$ is met ($\theta \approx 40^\circ$). For MQW device, due to the stronger interaction with free space radiation, the blueshift of the main plasmon peak increases with $\theta$ and critical coupling condition $\gamma=n_{QW}(\theta) \Gamma(\theta)$
is fulfilled at lower angle ($\theta \approx 20^\circ$). Beyond $\theta \approx 40^\circ$, the second plasmon peak at 185 meV becomes dominant. 
The comparison between experimental and theoretical results shows that our model provides a very good understanding of the variations of the linewidth, position and amplitude of the plasmon modes in both samples 
and supports their interpretation in terms of single photon superradiance and cooperative Lamb shift. 

The bottom panels of fig.~\ref{lamb} show the variations of the emitted power per solid angle with $\theta$ for the two devices. At very low angles, the emitted power increases proportionally to $\Gamma(\theta)$ for SQW (left) and to 
$n_{QW}(\theta) \Gamma(\theta)$ for MQW (right), as it would be expected from a perturbative treatment of the light-matter interaction. However, at higher angles light-matter interaction is non-perturbative and the two devices display different angular behaviors, with a maximum emission at 35$^\circ$ (55$^\circ$) for the MQW (SQW) sample. 
The predictions of our model (full lines) are well corroborated by the experimental results for the total power (colored dots). This further confirms that the observed behavior is a consequence of the superradiant enhancement 
of the radiative rate in MQW device. Note that although the two devices have identical electrical characteristics, the maximum emitted power is significantly increased in MQW with respect to SQW, showing that multiple QW 
superradiance is a possible approach to improve the performance of thermal emitters in the mid-infrared. 

In summary we demonstrated a room temperature semiconductor platform that allows probing some of the most fundamental properties of quantum 
electrodynamics, like superradiance and Lamb shift, which are usually the realm of atomic physics. Furthermore, the observed effects open new perspectives in the development of efficient mid-infrared sources. 

\section*{Funding Information}
Labex SEAM; Agence Nationale de la Recherche (grant ANR-14-CE26-0023-01); Renatech.

\bigskip \noindent See \href{link}{Supplement 1} for supporting content.

\bibliographystyle{apsrev4-1}
\bibliography{biblio}

\end{document}